# GLOBAL SYSTEMS PERFORMANCE ANALYSIS FOR MOBILE COMMUNICATIONS (GSM) USING CELLULAR NETWORK CODECS

MaphuthegoEtu Maditsi, Thulani Phakathi,
Francis Lugayizi and MichaelEsiefarienrhe

Department of Computer Science,
North West University, Mahikeng, South Africa

*ABSTRACT*

*Global System for Mobile Communications (GSM) is a cellular network that is popular and has been growing in recent years. It was developed to solve fragmentation issues of the first cellular system, and it addresses digital modulation methods, level of the network structure, and services. It is fundamental for organizations to become learning organizations to keep up with the technology changes for network services to be at a competitive level. A simulation analysisusing the NetSim tool in this paper is presented for comparing different cellular network codecsfor GSM network performance. Theseparameters such as throughput, delay, and jitter are analyzed for the quality of service provided by each network codec. Unicast application for the cellular network is modeled for different network scenarios. Depending on the evaluation and simulation, it was discovered that G.711, GSM_FR, and GSM-EFR performed better than the other codecs, and they are considered to be the best codecs for cellular networks.These codecs will be of best use to better the performance of the network in the near future.*

*KEYWORDS*

*GSM, CODECS, Cellular Network, G.711, GSM_FR, GSM-EFR.*

## 1. INTRODUCTION

Cellular Networks play an important role in most organizations in the Information Technology (IT) field and countries as it potentially improve the economic growth and it also contributes extremely to human development and employment. Mobile communication networks equally help in sharing and gathering information [1]. With that being said, cellular networks have become very fundamental in our daily lives in a way that we cannot live without them, hence, the number of mobile subscribers to networks increases on regular basis. The evolving technology such as the 5G mobile network, makes today's network monumental in size and complexity. The consumption of mobile network data is so high that it puts a lot of strain on the structure of the network due to the limited radio spectrum. This may lead to performance degradation that causes lower Quality of Service (QoS) or poor network performance [2]. Moreover, poor services to the customers include dropped calls, insufficient bandwidth, and slow response time for data downloads to mention but a few. Good quality of the network will retain customers whereas a poor network service will cause a high level of churn for the cellular network operator. It is the duties of the Network Management System (NMS) to make the network as efficient as possible. The NMS is composed of Fault Management, Configuration Management, Accounting Management, Performance Management, and Security Management (FCAPS) functional areas and as well as traffic prediction and congestion control [3]. There are various mobile communication standards such as GSM, Code Division Multiple Access (CDMA), Long Term Evolution (LTE), e.tc. The motive for various communication standards was introduced to improve technical performance leading to an acceptable level ofQoS in cellular networks/mobile communication networks [4]. GSM is the most reliable network coverage that has maximum capacity and confidentiality and it is a network used for transmitting mobile voice and data services. During this transmission of voice and data, the network may encounter various obstacles like phone echoes and loss of signals that may be caused by network latency, insufficient bandwidth, delay, jitter, destructive interference, congestion in the towers, and other external factors. This research work focuses on performance analysis of the Global System for Mobile Communications (GSM) network with the view to recommend the best cellular network codecs that will provide good performance and to determine features that need to be improved in the structure and operation of the network. The performance of the network will be analyzed through simulation where various codecs shall be compared and this is the basic building block for efficient transmission of data. The different codecs in the simulation to be used includes G.711, G.729, G.723, GSM-FR, and GSM-EFR. G.711 does not use compression at all and produce the best quality of call while G.729 at a low bit rate of 8kbps gives a good level of quality. This means one would be able to get more calls through bandwidth if you were to use the G.711 Codec [5]. GSM-FR is the first digital speech coding standard used in the GSM digital mobile phone system and operates at a bitrate of 13kbps [5]. GSM-EFR has better quality of speech and better robustness to network impairments [5]. G.723 is for speech and uses extensions of G.721 that provide voice quality using adaptive differential pulse code modulation [5].

This work is arranged as in the following: Section 2 explains the basic functions of a GSM network, Section 3 explains various methods for performance analysis, Section 4 presents the proposed methods, and section 5 presents the results.

## 2. THE GSM NETWORK

The GSM is a standard that was developed by European Telecommunication Standard Institute (ETSI) to describe various protocols for the second-generation digital cellular networks used by mobile devices such as mobile phones and tablets [6]. It was first deployed in Finland in December 1992 to meet certain criteria like support for a wide range of latest services, the security of transmission, compatibility with the fixed voice network, and the data networks are improved concerning the existing first-generation systems [7].

GSM as a network consists of Frequency Division Multiple Access(FDMA) and Time Division Multiple Access(TDMA) where FDMA is about dividing the frequency band into different ones and TDMA allows the same channel of frequency to be used by different subscribers [8]. This makes every user have their time slot to use a specific frequency allocated to them. GSM is different in that it utilizes TDMA approaches to subdividing carrier frequency that is further segmented into separate time slots. GSM composes of three major systems namely, the Switching System (SS), the Base Station System (BSS), and the Operation and Support System (OSS).

### 2.1. The Switching System

The importance of the switching system is to perform call processing and functions of the related subscribers. The following are the functional unit of these switching systems:

**Home Location Register (HLR)**, a database that is used to permanently store subscriber's data that includes a subscriber's service profile, location information, and activity status. The user is registered in the HLR of the operator that is buying a subscription from.

**Mobile Service Switching Centre (MSC)**, is responsible for performing telephony functions of the system like routing calls and other services like FAX and conference calls.

**Visitor Location Register (VLR)** is responsible for storing temporary information about subscribers that is required by the MSC to service visiting subscribers. When a mobile station roams into a new area of MSC, the VLR that is always integrated into that MSC will request data about the mobile station from the HLR. In case the mobile station makes a call then the VLR will have the information that is required for setting up the call without the need to request it from the HLR each time.

**Authentication Centre (AUC)** provides security for network users by providing authentication and encryption capabilities.

**Equipment Identity Register (EIR)**, stores information about the identity of mobile equipment that prevents calls from stolen, defective mobile stations.

## 2.2. The Base Station System (BSS)

The BSS consists of the Base Station Controller (BSC), Base Transceiver Station (BTS), and where all the functions related to radio are performed.The BSC is primarily responsible for all the control functions and physical links between the MSC and BTS.

## 2.3. The Operation and Support System (OSS)

All the equipment in the switching systems iconnected to the operation and maintenance center. The main responsibility of the OSS is to service customers with support that is cost-effective for all operations and maintenance activities in the the GSM network. Figure 1 shows the GSM network structure [7].

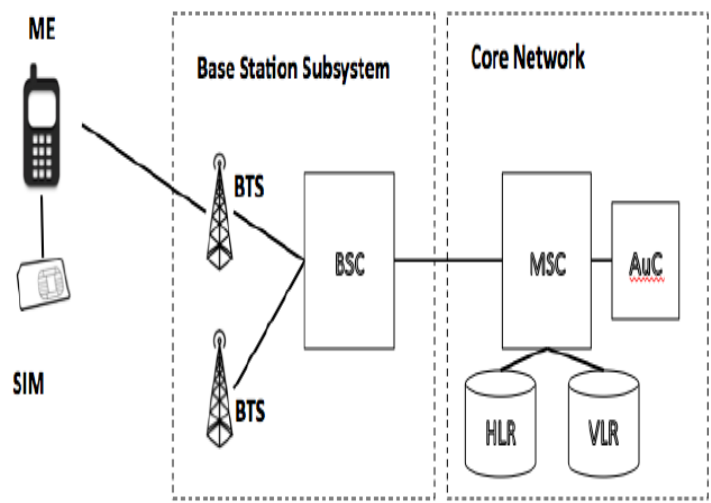

Figure 1.The basic structure of the GSM network

## 3. THE RELATED WORK

### 3.1. Literature Review

To established the research gap from literature and to better understand the efforts of the work studied, a systematic review was carried out.As according to authours in [8], it is the critical and verifiable summary arising from the various publications that address the aim of this research done by various researchers globally related to the field of study.

Several researchers has proffered various solutions to performance analysis of GSM using cellular network codecs that are equally robust and challemging. Here the review of literature is provided to gain insight into what has been achieved thus far and to indicate how our work differ from previous work. The authors in [6] proposed performance analysis based on GSM Key Performance Indicators (KPIs) where they described the importance of pre-selecting relevant KPIs to focus on when analyzing and monitoring the network performance. The authors as above proposed the use of generating measurement data from a live network with the sole aim of analyzing and evaluating the results for optimizing the performance of the GSM network. Their work was proposed to contribute to the Systematic examination of GSM mobile networks performance and end-user experience using four NCC KPIs metrics.

Moreover, the authors in [7] analyzed the performance of several audio encoding schemes using Resource Reservation Protocol (RSVP) based on Wireless Local Area Networks (WLAN). The codec G.711, G.723, and G.729 were compared to come up with results using the RSVP services. On the other-hand voice, codecs were used by the authors in [8] to examine the performance of mobile Worldwide Interoperability for Microwave Access (WiMAX). The simulation results showed that the performance of audio codecs stays fixed for random waypoint and group mobility manner while this is in contrast with our work because the performance of the codecs is not fixed for random ER_LANG call application of different nodes.

The most important KPIs from the author's view in [6] include Bit Error rate (BER), Frame Erasure Rate(FER), Bit Error Probability (BEP), and Mean Opinion Score (MOS). These are main KPIs that NCC used for rating the quality of service of cellular networks in Nigeria [8]. Wireless communication is expanding rapidly as most users are switching to it. The capacity of cells in the existing digital cellular mobile networks like GSM needs to meet the requirements that come with these changes in the network. The issue faced is how to increase the capacity of an existing network without causing performance or QoS degradation while our work focuses on a modeled network with increasing nodes that doesn't affect the service provided.

The authors in [9] proposed that the GSM band should be increased, along with increasing the number of serving channels or frequencies in an area. Although the overall spectrum of GSM is limited and it is divided between two or three network operators, leaving a spectrum of not more than 10 MHz for each operator. Another solution that was proposed by the author in [9] was to deploy more base stations or to introduce hierarchical structures like micro and pico-cells [10,11]. This method is limited to a denser base station grid resulting in increased interference. This limits the quality and capacity in terms of soft blocking.

Also, in literature, several analytical models based on continuous-time Markov chains have been proposed for studying the performance problems in GSM networks. The authors in [12] evaluated the impact of reserving channels for data and multimedia services on the circuit switch GSM network performance. Under a given GSM call characteristics,theauthours in [13] developed a model called the Markov model that was used to analyze the performance of GPRS. The authors in [14] also developed a Markov model to derive average data and restricting probabilities rates for GPRS in GSM networks.

In [15] the performance of a personal communication network based upon microcells wasanalyzed to find some fundamental phone traffic parameters such as blocking probabilities of new calls and handovers. A wireless system that has traffic scenarios based on Poisson time-dependent process is described using the fluid model [16]. The techniques to reduces dropped calls in progress due to failures of handovers are also proposed by the author in [17]. This approach describes several priority schemes, however, our work provides the use of codecs to maximize the QoS for cellular communications.

The authors in [18] analyzed is the performance of a hierarchical cellular system based on microcells and overlaying macrocells and also the advantage of introducing "tier handovers". These are handovers among cells that are from various hierarchical levels. As cellular networks collect a vast amount of measurement information that can be utilized to measure the performance of a network and QoS, the author in [19] studied the application of different data-analysis methods for processing the available measurement information. It is studied to produce more adequate methods for optimizing the performance while our study focuses on selecting the best performing codec through network simulation to maintain a better QoS.

The authors in [19] propose expert-based methods for monitoring and analyzing multivariable cellular network performance data. This method enables the analysis of performance bottlenecks that impacts performance indicators in multiple networks. Also, methods for more advanced failure diagnosis have been proposed to identify the causes of performance bottlenecks. The authors in [19] studied the use of measurement information in the identification of relevant optimization actions that leads to good network performance and good QoS, whereas our research focus is on analyzing performance measures to identify relevant audio codecs that give the best performance.

## 4. METHODS AND MATERIALS

The preceding research methods were applied in this work:

### 4.1. Simulation Process

Simulation is a process that models the behavior of a network by measuring the interaction between different network nodes [13]. The primary purpose of simulations is identifying problems that exist in a network or troubleshooting for unexpected interactions with one that has not been yet constructed. Simulation is mainly used in performance analysis, comparison, or even management and also for determining how a network would behave in a real-life situation. The generated results of the simulation aids in identifying the performance [12].

Discrete event simulation is a method used to model real-world structures, it is used to simulate the performance and behavior of a network [14]. The performance of codecs G.711, G.723, G.729, GSM-FR, and GSM-EFR is simulated where no change in the models were assumed during packets transmission [14]. This is done to address issues related to cellular traffic generated and handovers between neighboring calls, on a certain network capacity (number of BS, MSC, and cell towers). The network will be configured with the same properties including mobility model, type of protocol, application method, application type, QoS class, simulation duration, and various codecs. Different network codecs will be simulated to select the best one suitable for the networks. The codecs that produce a good QoS will be recommended for use in the network.

Table 1. Parameters

| Parameters | Configuration |
|---|---|
| Mobility Model | Random walk |
| Protocol | GSM |
| Application Method | Unicast |
| Application Type | ER_Lang Call |
| QoS service type | UGS |
| Codec(s) | G.711, G.723, G.729, GSM-FR and GSM-EFR |
| Simulation Duration | 100 s |
| Priority | High |

In this study, the QoS measures used to ensure that quality is optimized in the network are throughput, delay, and jitter. Performance measures refer to factors or parameters that are utilized to measure network performance. Each network varies in nature and design, therefore there is various way to measure it.

**Jitter**- these are voice packets that arrive at regular intervals to be intelligible. Jitter measures the degree of variation in packets interval which can be caused by improper queuing and configuration errors. The equation 1 below is used to calculate jitter where J is jitter, PJ is packet jitter, and N is the total amount of packets [15].

$$J = \frac{PJ}{(N-1)} \quad (1)$$

**Throughput**- network throughput is the rate at which voice packets are successfully transmitted over a network channel. It is the sum of all received packets by all nodes [15]. The equation 2 for throughput where S is the transmission time, T is the throughput, P is the average packet size and N is the total number of packet received, is as follows

$$T = \frac{(N*P)}{S} \quad (2)$$

**Delay**- delay in voice communications is defined as "the time it takes for voice packets to be transmitted from source to destination" which may lead to delay and echo [16]. Delay is measured in two ways, one direction, and a round trip. In one direction, the transmission of packets is unidirectional while in round trip voice packets travel to and from the destination and back to the source. Delay is measured in milliseconds (ms). The equation 3 shown below is used to compute delay where $\mu$ is the number of packets per second, D is the delay, and $\Lambda$ is the average rate at which packets arrive [17].

$$D = \frac{1}{\mu - \Lambda} \quad (3)$$

## 5. RESULTS AND DISCUSSION

**5.1.** In this chapter, the results of the network model simulation from the previous chapter are presented. The two different scenarios with different codecs will be compared. The quality of service is as well analyzed based on the measures, jitter, throughput, and delay. The best codec that provides good quality and performance will be recommended for cellular network usage. Each scenario will be fixed for unicast application and they will be analyzed based on QoS performance measures.

### 5.1.1.   G.711 And G.729

Figure 2 shows the throughput for application one that was generated during the G.711 simulation. As the time for transmitting the data increases, the throughput in the network also increases. The graph shows that close to 0.0114 Mbps voice is generated at 707 ms and after that remains constant. Figure 3 shows throughput results for G.729 simulation. Initially, there was a sharp increase in throughput that reached 0.0084 Mbps for the first 10000 ms, which gradually slows down from 20000 ms. As it, reaches 0.009 Mbps, the throughput starts to increase very slowly towards the end of the simulation.

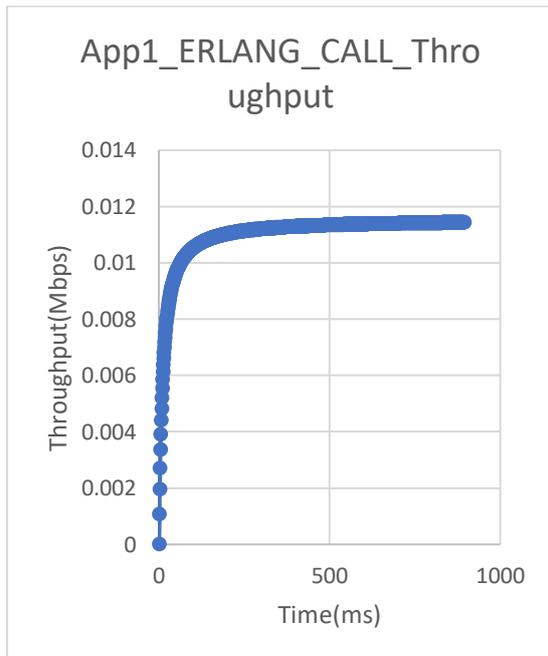
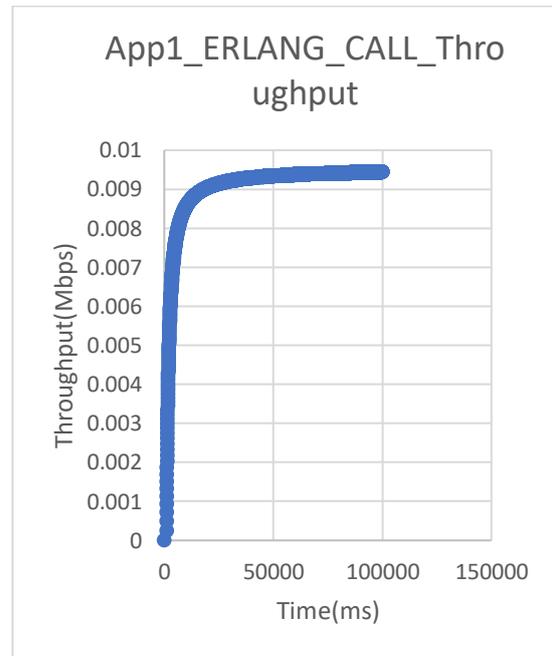

Figure 2 The throughput for G.711          Figure 3 The throughput for G.729

### 5.1.2.   G.723.gsm-fr and gsm-efr

In figure 4 the first 1000ms of simulation shows a sharp increase in throughput reaching 0.007 Mbps. The next 9000 ms shows a slow constant increase of 0.0079 Mbps until it reaches 0.007839 Mbps at the end of the simulation. Figure 5 illustrates the throughput for GSM-FR. Initially, there is a sharp increase in the throughput reaching 0.0055 Mbps. At the peak where simulation time was close to 19 990 ms, it begins to increase slowly from 0.007 Mbps until it reaches a constant throughput of 0.008 Mbps at 40000 ms of simulation time. More packets are transmitted in a lesser period, which results in good QoS, making codec G.723 desirable for use in a cellular network.The GSM-EFR figure 6 shows that throughput started to be transmitted at 0.0113 Mbps and begins to increase gradually with an increase in time from the beginning of the simulation.

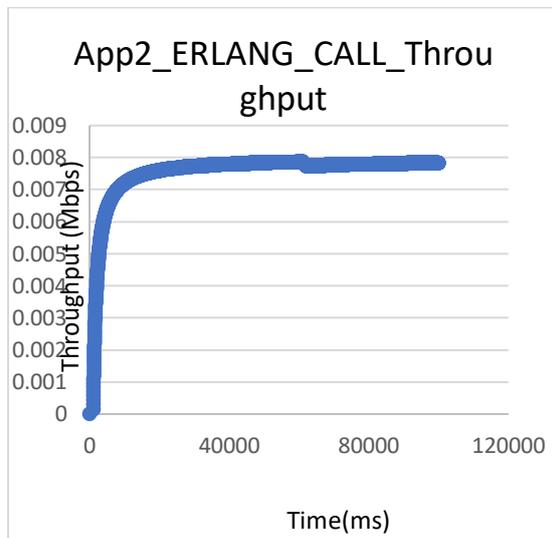

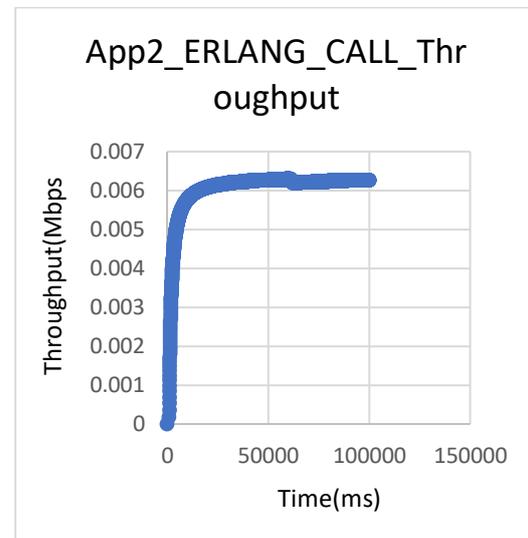

Figure 4. The throughput codec G.723          Figure 5. The throughput for GSM-FR

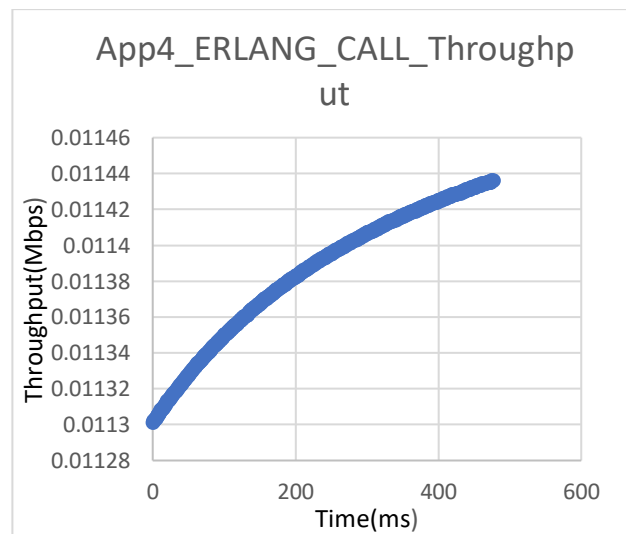

Figure 6. The throughput for GSM-EFR

### 5.2. 10-Nodes Scenario

A network model was created with 10 mobile stations. The results for throughput are shown in the scatter plot for each network codec.

#### 5.2.1. G.711 and G.729

The following Figure 7 with an increased number of nodes illustrates that Initially, there was no data sent until 60000 ms where the transmission started, and then after increase rapidly with an increase in time up until 0.0044 Mbps. More time is required to transmit data. Figure 8 illustrates a rapid increase in the throughput initially, and the time it takes to send data starts to increase after 0.003 Mbps. At 10000 ms,the achieved throughput is 0,008 Mbps. This gives a better performance of the codec.

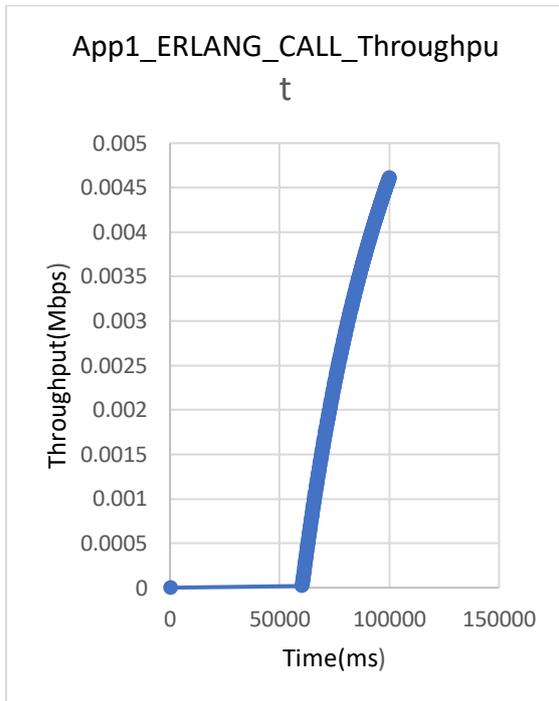 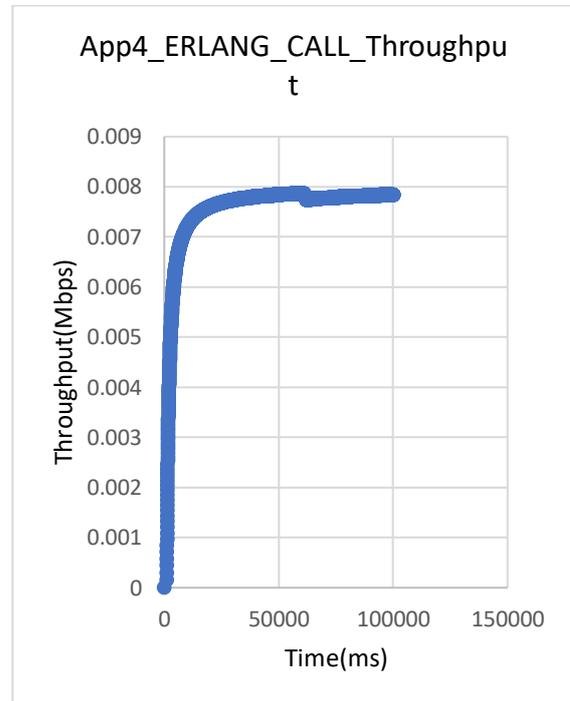

Figure 7. The throughput for GSM-EFR                    Figure 8. The throughput for G.729

### 5.2.2.  G.723, GSM-FR and GSM-EFR

In Figure 9 thw throughput for G.723 illustrates that the data send increases as time increases. There is a gradual increase in throughput initially. After 10000 ms the throughput was 0.006 Mbps and begins to increase very slowly at 20000 ms of simulation time. There is a slight decrease of 0.0061 Mbps in throughput just after 60000 ms and continues with a slow increase of 0.0062 Mbps for the rest of the simulation time. Figure 10 shows that the throughput increases fast at the beginning of the transmission and slows down a bit to 0.004 Mbps after the first 10000 ms of simulation time. It gradually reached a peak of 0.0087 Mbps. Figure 11 shows the throughput for codec GSM-EFR increases gradually until it reaches 0.008 Mbps after the first 10000 ms of simulation time. The throughput begins to slow down from 0.0084 Mbps as time increases. It starts to be constant at 60000 ms with a throughput of about 0.0088 Mbps.

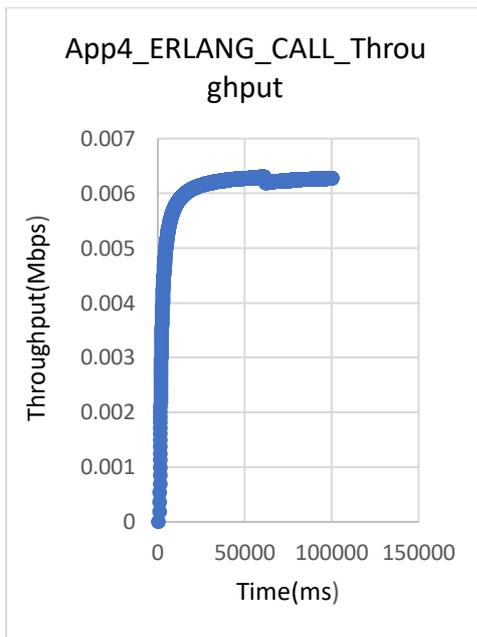

Figure 9. The throughput for G.723

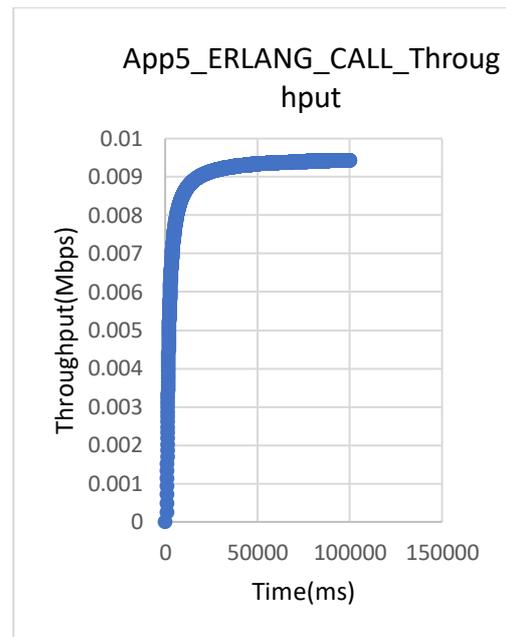

Figure 10. The throughput for GSM-FR

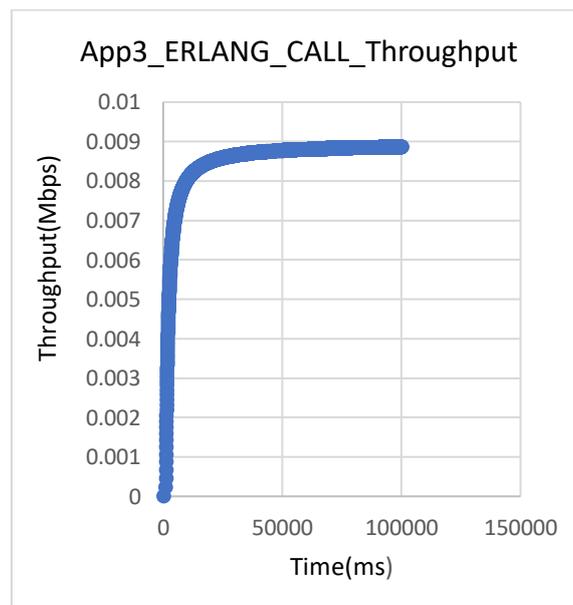

Figure 11. The throughput for GSM-EFR

## 5.3. Jitter and Delay

Jitter is the variation of the time between the packets that are arriving caused by the network congestion or route changes while throughput is the actual bandwidth that is measured with a specific time unit used to transfer data of a certain size [16].

### 5.3.1. G.711 AND G.729

The Figure 12 below for codec G.711 shows that there is a 31932.403879µs difference in delay for the 5-Nodes scenario and 10-Nodes scenario. According to the graph, there is no jitter experienced during packet transmission using this codec. Figure 13 for codec G.729 illustrates that the delay for 5-Node is less than that of the 10-Node scenario with 3.509242µs. The jitter increases with the increase in the number of nodes. The graph shows that the jitter for 5-Nodes is less than the jitter for the 10-Nodes scenario.

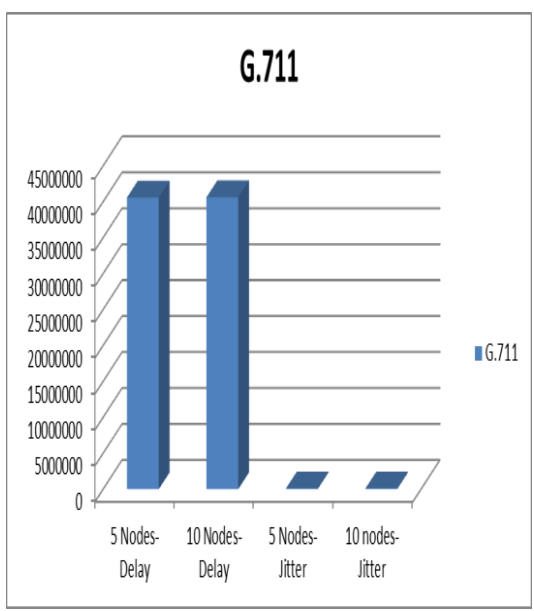 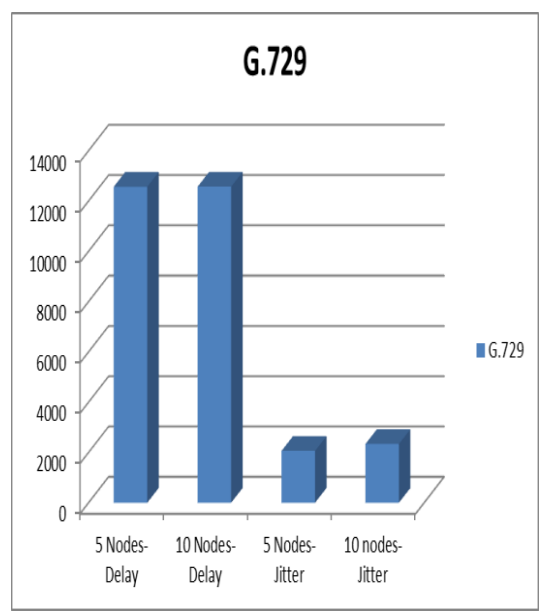

Figure 12. Variation in jitter and delay for G.711    Figure 13. Variation in jitter and delay for G.729

### 5.3.2. G.723, GSM-FR and GSM-EFR

The following Figure 14 clearly shows that the delay experienced during transmission of packets using the G.723 codec is the same for both scenarios. The number of nodes does not have an impact on delay. There is high jitter of 12569.741364µs for 5-nodes and 12566.232122 µs for 10-nodes connected to the network. Figure 15 for codec GSM-FR shows that for 5 nodes and 10 nodes, there is a delay of 14000000µs. There was no indication of jitter during the transmission of packets according to the results simulated for both scenarios. Desired data is transmitted but at an undesirable time.Figure 16 for codec GSM-EFR indicates that both scenarios have a delay of 14000000µs. During the transmission of voice, no data is lost and this leads to better QoS.

58    Computer Science & Information Technology (CS & IT)

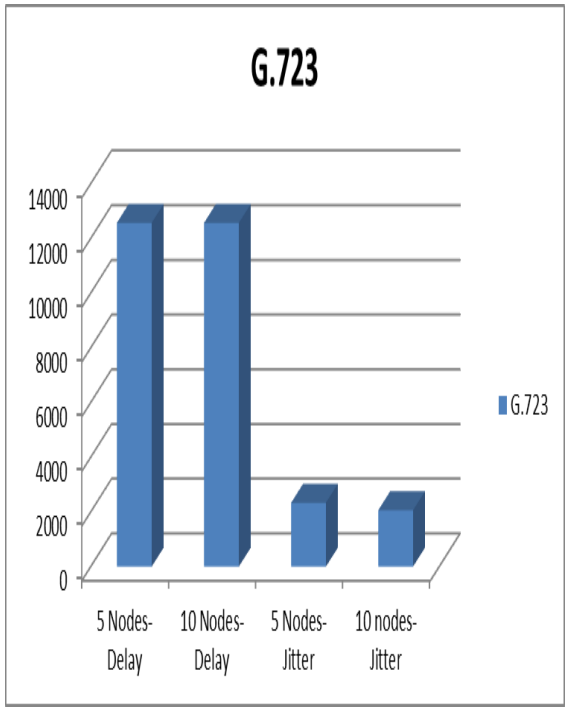

Figure 14. Variation in jitter and delay for G.723

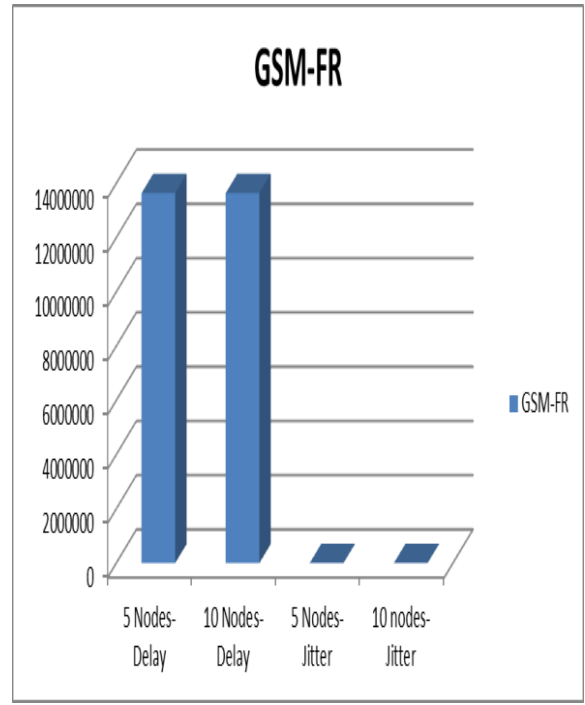

Figure 15. Variation in jitter and delay for GSM-FR

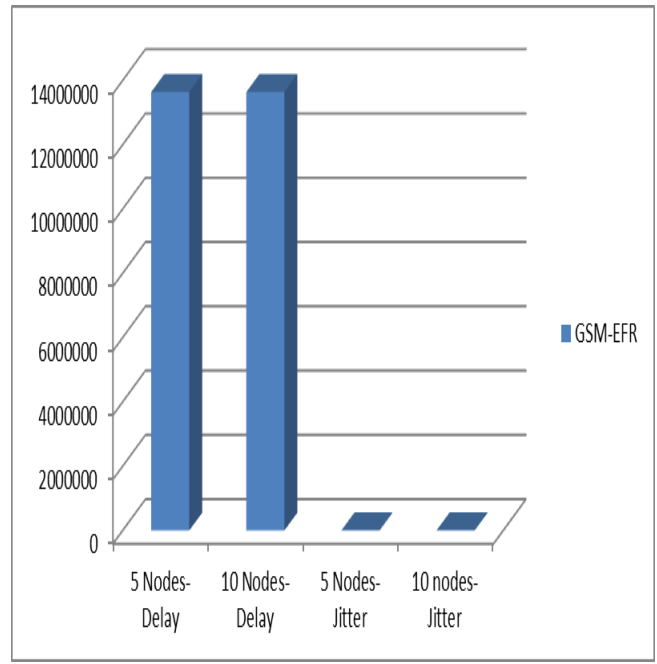

Figure 16. Variation in jitter and delay for GSM-EFR

## 5.4. Discussion

The variance in delay, jitter, and throughput among different codecs was simulated. G.729 operates differently compared to other codes. The lesser the nodes the higher the jitter and vice versa. More jitter is experienced for 10 nodes and less for 5 nodes. The delay that occurred in G.723 has a slight difference between the scenarios, it is almost the same and more jitter is experienced with fewer mobile stations. The codec G.711, GSM-FR, and GSM-EFR have a constant delay and does not experience any jitter for both scenarios, the number of mobile stations doesn't matter in this case. Therefore G.711, GSM-FR, and GSM-EFR will best suit the cellular network.

## 6. CONCLUSIONS

It can be concluded that the codec G.711, GSM-FR, and GSM-EFR showed the best performance for both 5-Nodes and 10-Nodes scenarios. G.723 and G.729 performed poorly for both scenarios. This was in terms of throughput, delay, and jitter.To maximize the value of QoS, it is very fundamental to appropriately utilize codecs in analyzing GSM network performance. The aim of this paper is shown from the simulation results that show a selection of G.711, GSM-FR, GSM-EFR as they produce significant results for the performance of the GSM cellular network. These codecs have no jitter and delay during transmission of data compared to other codecs like G.723, and G.729. This means that data packets will be sent in the desired period, and good quality of service will be provided to users using the selected codec for the network.As a results G.711, GSM-FR and GSM-EFR will positively impact the network performance.

## ACKNOWLEDGEMENTS

This research work was supported by the research entity MaSIM of the FNAS, North-West University and our partners at TETCOS,India.